\documentclass[conference, a4paper]{IEEEtran}

\usepackage{amsmath}
\usepackage{amsfonts,amssymb}
\usepackage{array}
\usepackage{graphicx}
\usepackage{booktabs}
\usepackage{afterpage}
\usepackage{float}
\usepackage{amsbsy}
\usepackage{mathtools}
\usepackage{color}

\def\bA{{\mathbf{A}}}  \def\bC{{\mathbf{C}}}  
 \def\bG{{\mathbf{G}}} \def\bH{{\mathbf{H}}}  
  \def\bM{{\mathbf{M}}}  
 \def\bQ{{\mathbf{Q}}}   
\def\bU{{\mathbf{U}}} \def\bV{{\mathbf{V}}} \def\bW{{\mathbf{W}}} \def\bX{{\mathbf{X}}} \def\bY{{\mathbf{Y}}}
\def\bZ{{\mathbf{Z}}}
    
\def\bf{{\mathbf{f}}} \def\bg{{\mathbf{g}}}

\def\C{{\mathbb{C}}}

\begin{document}
\title{Channel  Estimation for Hybrid RIS Aided  MIMO Communications via Atomic Norm Minimization}
\author{\IEEEauthorblockN{Rafaela Schroeder\IEEEauthorrefmark{1},
Jiguang He\IEEEauthorrefmark{1}, and Markku Juntti\IEEEauthorrefmark{1}}
\IEEEauthorblockA{\IEEEauthorrefmark{1}Centre for Wireless Communications, FI-90014, University of Oulu, Finland} 
\IEEEauthorblockA{E-mail:\{rafaela.schroeder, jiguang.he, markku.juntti\}@oulu.fi}}
\maketitle
\begin{abstract}
Reconfigurable intelligent surfaces (RISs) have been introduced as a remedy for mitigating frequent blockages in millimeter wave (mmWave) multiple-input multiple-output (MIMO) communication networks. However, perfect or nearly perfect channel state information (CSI) is fundamental in order to achieve their full potential. Traditionally, an RIS is fully passive without any baseband processing capabilities, which poses great challenges for CSI acquisition. Thus, we focus on the hybrid RIS architecture, where a small portion of RIS elements are active and able to processing the received pilot signals for estimating the corresponding channel. The channel estimation (CE) is done by resorting to off-the-grid compressive sensing technique, i.e., atomic norm minimization, for exacting channel parameters through two stages. Simulation results show that the proposed scheme outperforms the passive RIS CE under the same training overhead.     
\end{abstract}

\section{Introduction}
Millimeter-wave (mmWave) multiple-input multiple-output (MIMO) communications is a promising technology to address the ever-increasing demand for high data rate in 5G and beyond~\cite{alkhateebCEHybridprecod,heath2016overview, rappaportmmWave}. Large antenna arrays are essential at both the transmitter and receiver to compensate for the severe path loss. However, difficulty on channel state information (CSI) acquisition surges. It is well known that mmWave channel is more susceptible to line-of-sight (LoS) blockage~\cite{heath2016overview}, and mitigating this can be well addressed by deploying a  
reconfigurable intelligent surface (RIS) to the network. 
The RIS consists of a large number of elements that can be 
controlled to modify its signal response in order to achieve a certain objective, for example, reflecting the signals towards the receiver~\cite{ardah2020trice}. Namely, RIS enables the control of the propagation environment, which is infeasible in the past. The RIS has been verified as an innovative technology to further improve the performance of mmWave MIMO systems, in terms of indoor localization~\cite{ma2020indoor}, security~\cite{yang2020secrecy}, and spectral efficiency (SE)~\cite{wu2019intelligent,zhang2020capacity}.

The efficiency and accuracy of CSI acquisition plays an important role in design of the RIS phase control matrix and active beamforming vectors at the transceivers. Compared to mmWave systems, CSI acquisition becomes more challenging in RIS-aided ones due to increased complexity. In the literature, CSI acquisition has been investigated for both RIS architectures, i.e., \textit{passive RIS} and \textit{hybrid RIS}. The architecture with both passive and \emph{active} elements is called \textit{hybrid} RIS.
The channel estimation (CE) methods for passive RIS-aided mmWave MIMO systems have been intensively studied in~\cite{ardah2020trice,he2020anm,CE_He_Zhen-Qing,CE_mirza2019,he2020,de2020parafac}. Ardah~\textit{et al.}~\cite{ardah2020trice} proposed a two-stage procedure for the CE for mmWave MIMO systems. The CE problem was solved by following the parallel factor (PARAFAC) tensor decomposition in~\cite{de2020parafac}. In our recent work~\cite{he2020anm}, we also considered two-stage CE with the aid of atomic norm minimization (ANM). 
In order to ease the CE procedure, Taha \textit{et al.}~\cite{ActElements} introduced a few \emph{active} elements at the RIS with baseband capabilities, which allow the CE at the RIS despite the higher power consumption. Inspired by~\cite{ActElements}, the hybrid RIS architecture has been further studied in~\cite{oneRFchain, schroeder2020passive,nguyen2021hybrid,lin2021tensor}. In~\cite{schroeder2020passive}, we made a comparison in terms of CE between the passive and hybrid architectures for mmWave MIMO systems, which concludes that passive RIS is superior to the hybrid one when CE is done at the RIS. 

In this paper, we develop a new two-stage CE method under the hybrid RIS architecture. Unlike our previous work~\cite{schroeder2020passive}, the proposed CE requires fewer number of active RIS elements and only one-way training (either uplink or downlink). We consider uplink training, where the mobile station (MS) sends pilots to the base station (BS) via the RIS. The pilot signal is also received at the RIS by the active elements, while the remaining  elements reflect the signal towards the BS. In the first CE stage, we aim at recovering the parameters of the  MS-RIS channel via ANM based on the received signal at the RIS. 
In the second CE stage, we aim at recovering the parameters of RIS-BS channel from the received signal at the BS.
We evaluate the performance in terms of the mean square error (MSE) of the  angle of departure (AoD) angle of arrival (AoA), and the products of path gains, and SE. In the evaluation, we also incorporate the path loss model, which has a nonnegligible  impact on CE. It is verified that our proposed method can simplify the CSI acquisition for RIS aided mmWave MIMO systems with a small number of active RIS elements and bring better CE performance compared to that in the passive RIS~\cite{he2020anm}.

\textit{Notation}: A bold capital letter $\mathbf{A}$ denotes a matrix and a lowercase letter $\mathbf{a}$ denotes a column vector, and $()^{\mathsf{H}}$, $()^*$, and $()^{\mathsf{T}}$ denote the Hermitian transpose, conjugate, and transpose, respectively. $\otimes$ denotes the Kronecker product, $\odot$ is the Khatri-Rao product, $\mathrm{vec} (\mathbf{A})$ is the vectorization of $\mathbf{A}$, $\mathrm{diag}(\mathbf{a})$ being a square diagonal matrix with entries of $\mathbf{a}$ on its diagonal, $\|.\|_{\mathrm{F}}$ is the Frobenius norm, $\mathrm{Tr} (\mathbf{A})$ is the sum value of the diagonal elements of $\mathbf{A}$, $\mathbb{T}(\mathbf{A})$ denotes the block Toeplitz matrix constructed from the vectorized form of $\mathbf{A}$, i.e., $\mathrm{vec}(\mathbf{A})$, being its first row. $(\cdot){\dagger}$ denotes the Moore–Penrose inverse, $\mathcal{A}$ is the atomic set, $\mathrm{conv}({\mathcal{A}})$ denotes the convex hull of $\mathcal{A}$, and $\mathbb{E}$ is the expectation operator. 

\section{System Model}
\label{sec: system_model}
The RIS-aided mmWave MIMO system consists of one multi-antenna BS, one multi-element RIS, and one multi-antenna MS. The number of antennas or reflecting elements at the BS, RIS, and MS are denoted by $N_\text{B}$, $N_\text{R}$  and $N_\text{M}$, respectively. We assume that the direct MS-BS channel suffers from blockage, which motivates the use of the RIS to assist the data transmission between the BS and the MS. We adopt an uniform linear array (ULA) for the antennas/elements in this paper, while it is possible to be extended for an uniform planar array (UPA). Regarding CE, we consider uplink training with pilot transmission from the MS. 
\subsection{Channel Model}
The propagation channel is composed of two tandem channels, i.e., MS-RIS and RIS-BS channels, denoted by $\bH_{\text{M,R}}\in\C^{N_{\text{R}}\times N_{\text{M}}}$ and $\bH_{\text{R,B}}\in\C^{N_{\text{B}}\times N_{\text{R}}}$, respectively. We adopt a block-fading channel, which means that the channel parameters stay constant during the coherence time. By adopting the geometric channel model, we define $\bH_{\text{M,R}}$ as
\begin{align}
\label{eq:hrm}
    \bH_{\text{M,R}} &= \sum\limits_{l = 1}^{L_{\text{M,R}}} [\boldsymbol{\rho}_{\text{M,R}}]_l \boldsymbol {\alpha}([\boldsymbol{\phi}_{\text{M,R}}]_l ) \boldsymbol{\alpha}^{\mathsf{H}}([\boldsymbol{\theta}_{\text{M,R}}]_l ),\nonumber\\
    &=\bA(\boldsymbol{\phi}_{\text{M,R}})\mathrm{diag}(\boldsymbol{\rho}_{\text{M,R}})\bA^{\mathsf{H}}(\boldsymbol{\theta}_{\text{M,R}}),
\end{align}
where $[\boldsymbol{\rho}_{\text{M,R}}]_l$ is the $l$th propagation path gain, $L_{\text{M,R}}$ is the number of paths, $\boldsymbol{\theta}_{\text{M,R}}$ and $\boldsymbol{\phi}_{\text{M,R}}$ are the AoDs and AoAs of the channel, respectively. We define $\boldsymbol{\alpha}([\boldsymbol{\phi}_{\text{M,R}}]_l )$ and $\boldsymbol{\alpha}([\boldsymbol{\theta}_{\text{M,R}}]_l)$ are the array response vectors as a function of $[\boldsymbol{\phi}_{\text{M,R}}]_l$ and $[\boldsymbol{\theta}_{\text{M,R}}]_l$. Considering half-wavelength inter-antenna element spacing, the array response vectors are defined as $[\boldsymbol{ \alpha}([\boldsymbol{\phi}_{\text{M,R}}]_l )]_{n} =  \exp\{
j\pi(n-1)\sin( [\boldsymbol{\phi}_{\text{M,R}}]_l)\}$, for $n=1,\cdots,N_\text{R}$ and $[\boldsymbol{\alpha}([\boldsymbol{\theta}_{\text{M,R}}]_l)]_{n} =  \exp\{j\pi(n-1)\sin( [\boldsymbol{\theta}_{\text{M,R}}]_l)\}$, for $n=1,\cdots,N_\text{M}$ and $j = \sqrt{-1}$. The array response matrices $\bA(\boldsymbol{\theta}_{\text{M,R}})$ and $\bA(\boldsymbol{\phi}_{\text{M,R}})$ are defined as 
\begin{equation}
    \bA(\boldsymbol{\theta}_{\text{M,R}}) = [\boldsymbol{\alpha}([\boldsymbol{\theta}_{\text{M,R}}]_{1}), ..., \boldsymbol{\alpha}([\boldsymbol{\phi}_{\text{M,R}}]_{L_{\text{M,R}}})],
\end{equation}
\begin{equation}
    \bA(\boldsymbol{\phi}_{\text{M,R}}) = [ \boldsymbol{\alpha}([\boldsymbol{\phi}_{\text{M,R}}]_{1}), ..., \boldsymbol{\alpha}([\boldsymbol{\phi}_{\text{M,R}}]_{L_{\text{M,R}}})].
\end{equation}
Similarly, the RIS-BS channel, $\bH_{\text{R,B}}$, is given by
\begin{align}
    \label{eq:hbr}
    \bH_{\text{R,B}} &= \sum\limits_{l = 1}^{L_{\text{R,B}}} [\boldsymbol{\rho}_{\text{R,B}}]_l \boldsymbol{ \alpha}([\boldsymbol{\phi}_{\text{R,B}}]_l ) \boldsymbol{\alpha}^{\mathsf{H}}([\boldsymbol{\theta}_{\text{R,B}}]_l),\nonumber \\
    & = \bA(\boldsymbol{\phi}_{\text{R,B}})\mathrm{diag}(\boldsymbol{\rho}_{\text{R,B}})\bA^{\mathsf{H}}(\boldsymbol{\theta}_{\text{R,B}}),
\end{align}
where $[\boldsymbol{\rho}_{\text{R,B}}]_l$, $\boldsymbol {\alpha}([\boldsymbol{\phi}_{\text{R,B}}]_l )$, and $\boldsymbol{\alpha}([\boldsymbol{\theta}_{\text{R,B}}]_l)$ are the $l$th propagation path gain, and array response vectors as a function of $[\boldsymbol{\phi}_{\text{R,B}}]_l$ and $[\boldsymbol{\theta}_{\text{R,B}}]_l$, respectively. The entire MS-RIS-BS channel based on~\eqref{eq:hrm} and~\eqref{eq:hbr} is expressed as 
\begin{equation}
\label{eq:hc}
      \bH= \bH_{\text{R,B}}\boldsymbol{\Omega}\bH_{\text{M,R}},
\end{equation}
where $\boldsymbol{\Omega}\in\mathbb{C}^{N_{\text{R}}\times N_{\text{R}}}$ is the phase control matrix at the RIS with constant-modules entries on the diagonal. We assume that the RIS is composed of discrete phase sifters. Thus, the phase control matrix is $\left[\mathbf{\Omega}\right]_{k,k}=\exp\left({j\omega_k}\right)$, where $\omega \in [0,2\pi)$. We also define the effective channel $\bG\in \mathbb{C}^{L_{\text{R,B}} \times L_{\text{M,R}}}$ as 
\begin{equation}
\label{eq:g}
    \bG  = \mathrm{diag} (\boldsymbol{\rho}_{\text{R,B}})\bA^ {\mathsf{H}}(\boldsymbol{\theta}_{\text{R,B}}){\boldsymbol{\Omega}}\bA(\boldsymbol{\phi}_{\text{M,R}})\mathrm{diag}(\boldsymbol{\rho}_{\text{M,R}}),     
\end{equation}
which depends on the angle differences associated with the RIS, products of path gains, and the RIS phase control matrix. 

\subsection{Passive RIS}
The passive RIS without any baseband capabilities has been explored in the literature~\cite{he2020,ardah2020trice,CE_He_Zhen-Qing}. CE for the passive RIS is more challenging because the cascaded channels in~\eqref{eq:hc} can only be estimated at the MS (downlink) or BS (uplink). We target at the recovery of the channel parameters in~\eqref{eq:hrm} and~\eqref{eq:hbr}. With the assumption of block-fading channel, we divide one coherence time into two sub-intervals: one for CE and the other for data transmission. Moreover, the CE sub-interval is further divided into $K$ blocks, and within $k$th block MS sends a series of training sequences i.e., column vectors of $\bX_k$, for $k=1,\cdots,K$ to the BS. The pilot signal is first reflected at the RIS by $\mathbf{\Omega}_k$, and further combined at the BS by $\bW_k$. The received signal ${\bY_{\text{P}}}_{k}$ at the BS is summarized as follows: 
\begin{equation}
     \label{eq_y1}
    {\bY_{\text{P}}}_{k}=\beta_{\text{2}}\bW_{k}^{\mathsf{H}}\bH_{\text{R,B}}\boldsymbol{\Omega}_k\bH_{\text{M,R}}\bX_{k}+ \bW_{k}^{\mathsf{H}}\bZ_{k},          
    \mbox{for } k=1,\cdots,K,
\end{equation}
where $\bZ_k$ is the additive Gaussian noise with each entry distributed as $\mathcal{CN} (0,\sigma^2)$ and $\beta_{\text{2}} = \sqrt{\frac{1}{\beta(d_{\text{1}}, d_{\text{2}})}}$, where $\beta(d_{\text{1}}, d_{\text{2}})$ denotes the overall path loss of the BS-RIS-MS link with $d_{\text{1}}$ and $d_{\text{2}}$ being the distance between the MS and RIS and that between the RIS and BS, respectively.

\subsection{Hybrid RIS}
\begin{figure} [!t]
    \centering
    \includegraphics[width=\columnwidth]{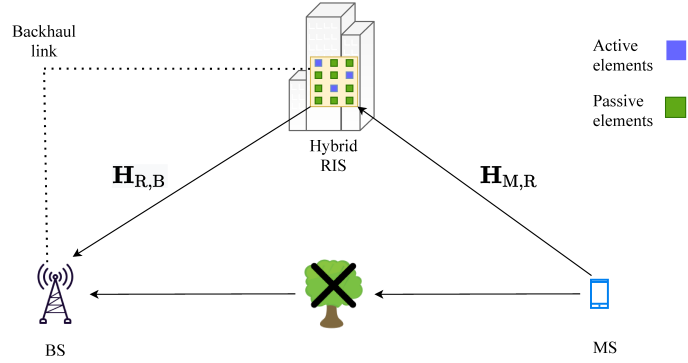}
    \caption{Hybrid RIS-aided communication system.}
    \label{fig1:RIS}
    \vspace{-0.6cm}
\end{figure}

The hybrid RIS with baseband capabilities simplifies the CE procedure for RIS-aided mmWave MIMO systems~\cite{ActElements}. The hybrid RIS enabled communication system is illustrated in the Fig.~\ref{fig1:RIS}. Thanks to the availability of baseband capabilities, the two tandem channels can be observed and estimated individually at the RIS. 

We consider $M$ out of $N_{\text{R}}$ RIS elements are active and $N_{\text{RF,R}} \leq M$ RF chains are implemented at the RIS. We assume uplink training, where the MS sends the training matrix $\bX\in\mathbb{C}^{N_{\text{M}}\times T}$ to the BS via the RIS with $T$ being the number of training beams. 
We further divide the CE sub-interval into $K$ blocks, each block with $T$ channel uses.

\begin{figure} [!t]
    \centering
    \includegraphics[width=\columnwidth]{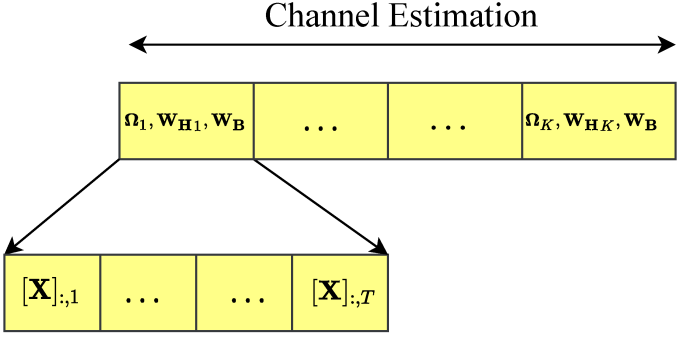}
    \caption{Uplink training procedure for the hybrid RIS.}
    \label{fig:training}
    \vspace{-0.6cm}
\end{figure}

The pilot signal is received at RIS by the $M$ active RIS elements indexed by set $\mathbb{A}$, i.e., $|\mathbb{A}| = M$, while the passive RIS elements reflect their received signals with a phase control matrix $\boldsymbol{\Omega}_{k}$. Note that here $\boldsymbol{\Omega}_{k}$ is different from that in passive RIS since $[\boldsymbol{\Omega}_{k}]_{i,i} = 0$ for $i \in \mathbb{A}$.  After reflection, the signal is further combined at the BS by $\mathbf{W}_{\text{B}} \in\mathbb{C}^{N_{\text{B}}\times N_{\text{C,B}}}$.
In the training procedure, we keep the combining matrix $\mathbf{W}_{\text{B}}$ and the training matrix $\bX$ constant over all the blocks. However, the phase control matrix $\boldsymbol{\Omega}_{k}$ varies from block to block. 
Fig.~\ref{fig:training} sketches the proposed uplink training procedure with a special focus on usage of training matrix, combining  matrix, and RIS phase control matrix. The received signal at the hybrid RIS is summarized as
\begin{equation}
\label{eq:r_s_ris_block}
      {\bY_{\text{H}}}_{k} = \beta_{\text{1}}{\bW_{\text{H}}}_k\bH_{\text{M,R}}\bX+{\bW_{\text{H}}}_k{\bZ_{1}}_k, \mbox{for } k=1,\cdots,K,
\end{equation}
where ${\bW_{\text{H}}}_k$ is the row-selection matrix containing $M$ rows of a $N_{\text{R}} \times N_{\text{R}}$ identity matrix. In the case of $N_{\text{RF,R}} < M$, further left-multiplying ${\bW_{\text{H}}}_k$ by a combining  matrix  $\bM_{\text{c}} \in \mathbb{C}^{N_{\text{RF,R}}\times M }$ is required, ${\bZ_{1}}_{k}\in \mathbb{C}^{N_{\text{R}} \times T}$ is the Gaussian noise with each entry distributed as $\mathcal{CN} (0,\sigma^2)$, and $\beta_{\text{1}} =\sqrt{\frac{1}{\beta(d_{\text{1}})}}$ with $\beta(d_{\text{1}})$ being the path loss for the MS-RIS channel. The received signal at BS is expressed as
\begin{equation}
    \label{eq:r_s_bs_block}
    \bY_k = \beta_{\text{2}}\bW_{\text{B}}^{\mathsf{H}}\bH_{\text{R,B}}\boldsymbol{\Omega}_k\bH_{\text{M,R}}\bX+ \bW_{\text{B}}^{\mathsf{H}}{\bZ_{2}}_k,
\end{equation}
where ${\bZ_{2}}_{k}\in \mathbb{C}^{N_{\text{B}}\times T}$ is the Gaussian noise at the BS. We collect the received signals across all the $K$ blocks at the RIS, as $[{\bY_{\text{H}}}_1^{\mathsf{T}},\cdots,{\bY_{\text{H}}}_K^{\mathsf{T}}]^{\mathsf{T}}
\in \mathbb{C}^{M K\times T}$, which can be summarized as 
\begin{equation}
        \label{eq:rris}
        \bY_{\text{H}} = \beta_{\text{1}}\bW_{\text{H}}\bH_{\text{M,R}}\bX+ \bar{\bZ}_{1},
\end{equation}
where $\bW_{\text{H}} = [{\bW_{\text{H}}}_1^{\mathsf{T}},\cdots,{\bW_{\text{H}}}_K^{\mathsf{T}}]^{\mathsf{T}}\in \mathbb{C}^{M K\times N_{\text{R}}}$, $\bar{\bZ}_{1} = [({\bW_{\text{H}}}_1{\bZ_{1}}_1)^{\mathsf{T}},\cdots,({\bW_{\text{H}}}_K{\bZ_{1}}_K)^{\mathsf{T}}]^{\mathsf{T}}\in \mathbb{C}^{M K \times T}$. 
The collected received signals at the BS $\bY = [\bY_1,\cdots,\bY_K]\in \mathbb{C}^{N_{\text{C,B}}\times T K}$ are expressed as 
\begin{equation}
 \label{eq:bs}
        \bY = \beta_{\text{2}}\bW_{\text{B}}^{\mathsf{H}}\bH_{\text{R,B}}\bU+ \bar{\bZ}_{2}, 
\end{equation}
where $\bU = [\boldsymbol{\Omega}_{1}\bH_{\text{M,R}}\bX,\cdots,\boldsymbol{\Omega}_{K}\bH_{\text{M,R}}\bX] \in \mathbb{C}^{N_{\text{R}}\times T K}$ and $\bar{\bZ}_{2}= [\bW_{\text{B}}^{\mathsf{H}}{\bZ_{2}}_1,\cdots,\bW_{\text{B}}^{\mathsf{H}}{\bZ_{2}}_K]\in \mathbb{C}^{N_{\text{C,B}}\times T K }$.

We propose a two-stage CE to recover the channel parameters from the received signals in~\eqref{eq:rris} and~\eqref{eq:bs}. To be specific, in the first stage, we extract the channel parameters in the $\bH_{\text{M,R}}$ from the received signals at the RIS~\eqref{eq:rris}. After that, the RIS sends the estimates of the channel parameters to the BS via the error-free backhaul link, and the BS reconstructs $\bH_{\text{M,R}}$ based on the estimates with the reconstructed one denoted as $\hat{\bH}_{\text{M,R}}$. In the second stage, we recover the channel parameters in $\bH_{\text{R,B}}$ from the received signals at the BS~\eqref{eq:bs} by assuming $\bH_{\text{M,R}} = \hat{\bH}_{\text{M,R}}$. More details will be provided in the sequel. 

\section{Hybrid RIS Channel Estimation}
\label{section: CE}
The CE for mmWave MIMO systems has been addressed in the literature via the off-the-grid  compressive sensing (CS) methods, such as ANM. Motivated by the recent works on off-the-grid methods~\cite{he2020anm,harnessing}, we also apply  ANM for hybrid RIS CE. 

\subsection{Preliminaries}
We apply the ANM for the estimation of the angles, or equivalent spatial frequencies. By replacing the angular parameters with spatial frequencies, we can express~\eqref{eq:hrm} and~\eqref{eq:hbr}, respectively,  as 
\begin{equation}
\bH_{\text{M,R}}= \sum\limits_{l = 1}^{L_{\text{M,R}}} [\boldsymbol{\rho}_{\text{M,R}}]_l \boldsymbol{\alpha}([\mathbf{g}_{1}]_l ) \boldsymbol{\alpha}^{\mathsf{H}}([\mathbf{f}_{1}]_l),
\end{equation}
\begin{equation}
\bH_{\text{R,B}}= \sum\limits_{l = 1}^{L_{\text{R,B}}} [\boldsymbol{\rho}_{\text{R,B}}]_l \boldsymbol{\alpha}([\mathbf{g}_{2}]_l ) \boldsymbol{\alpha}^{\mathsf{H}}([\mathbf{f}_{2}]_l),
\end{equation}
where $\bg_1 = \sin(\boldsymbol{\phi}_{\text{M,R}})$, $\bf_1 = \sin(\boldsymbol{\theta}_{\text{M,R}})$, $\bg_2 = \sin(\boldsymbol{\phi}_{\text{R,B}})$, and $\bf_2 = \sin(\boldsymbol{\theta}_{\text{R,B}})$. The spatial frequencies are within $[0,1)$, and the resultant atomic set of $\bH_{\text{M,R}}$ is denoted by $\mathcal{A}_{\text{M}}$, as 
\begin{equation}
    \mathcal{A}_{\text{M}} = \left \{ \bQ_{1}(\bf_{1},\bg_{1}): \bf_{1}\ \in \Big[ 0,1 \Big),  \bg_{1}\in \Big[ 0,1 \Big) \right \},
\end{equation}
where $\bQ_{1}(\bf_{1},\bg_{1}) = \boldsymbol{\alpha}(\bg_{1}) \boldsymbol{\alpha}^{\mathsf{H}}(\bf_{1})$ is the matrix atom. Similarly, the atomic set of $\bH_{\text{R,B}}$ is given by 
\begin{equation}
    \mathcal{A}_{\text{N}} = \left \{ \bQ_{2}(\bf_{2},\bg_{2}): \bf_{2}\ \in \Big[ 0,1 \Big),  \bg_{2}\in \Big[ 0,1 \Big) \right \},
\end{equation}
where $\bQ_{2}(\bf_{2},\bg_{2}) = \boldsymbol{\alpha}(\bg_{2}) \boldsymbol{\alpha}^{\mathsf{H}}(\bf_{2})$ denotes the matrix atom.

\subsection{Stage 1 CE}
To recover the matrix $\bH_{\text{M,R}}$, we formulate the atomic norm with respect to the atomic set $\mathcal{A}_{\text{M}}$ as
\begin{equation}
    \|\bH_{\text{M,R}}\|_{\mathcal{A}_{\text{M}}} = \inf \{q: \bH_{\text{M,R}} \in \mathrm{conv}({\mathcal{A}_{\text{M}}}) \} .
\end{equation}
The equivalent form as a semidefinite programming (SDP) problem is
\begin{equation}
    \nonumber
    \|\bH_{\text{M,R}}\|_{\mathcal{A}_{\text{M}}} = \mathrm{inf}_{\{\mathbf{C},\mathbf{V}\}} \Big\{\frac{1}{2 N_\text{M}} \mathrm{Tr}(\mathbb{T}(\mathbf{C})) + \frac{1}{2 N_\text{R}} \mathrm{Tr}(\mathbb{T}(\mathbf{V})) \Big\}
\end{equation}
\begin{center}
    \begin{equation}
        \textrm{s.t}   
        \begin{bmatrix}
        \mathbb{T}(\mathbf{C}) & \bH_{\text{M,R}}\\
        \bH_{\text{M,R}}^{\mathsf{H}} & \mathbb{T}(\mathbf{V})
    \end{bmatrix}
    \succeq 0.
    \end{equation}
\end{center}
where $\mathbb{T}(\bC)$ and $\mathbb{T}(\bV)$ are 2-level Toeplitz matrices. We can recover the angles $\boldsymbol{\theta}_{\text{M,R}}$ and $\boldsymbol{\phi}_{\text{M,R}}$ by addressing the following convex problem
\begin{equation}
    \label{eq:anm_h1}
    \hat{\bH}_{\text{M,R}}= \arg \min_{\bH_{\text{M,R}}} \tau \|\bH_{\text{M,R}}\|_{\mathcal{A}_{\text{M}}} + \frac{1}{2} \| \beta_{\text{1}}\mathbf{W}_{\text{H}}\bH_{\text{M,R}}\bX -  \bY_{\text{H}}\|^{2}_{\mathrm{F}}
\end{equation}
where $\tau$ is the regularization parameter set as $\tau \varpropto \sigma\sqrt{N_\text{R} N_\text{M}\log (N_\text{R} N_\text{M}})$. Using the SDP formulation, the problem can be expressed as 
\begin{equation}
            \nonumber
          \hat{\bH}_{\text{M,R}} = \arg \min_{\bH_{\text{M,R}},\bC,\bV} \frac{\tau}{2 N_\text{M}} \mathrm{Tr}(\mathbb{T}(\mathbf{C})) +  \frac{\tau}{2 N_\text{R}} \mathrm{Tr}(\mathbb{T}(\mathbf{V})) 
\end{equation}
\begin{equation}
        \nonumber
        + \frac{1}{2} \|\beta_{\text{1}} \mathbf{W}_{\text{H}}\bH_{\text{M,R}}\bX - \bY_{\text{H}}\|^{2}_{\mathrm{F}}
\end{equation}
\begin{equation}
    \textrm{s.t} \begin{bmatrix}
    \mathbb{T}(\mathbf{C}) & \mathbf{H}_{\text{M,R}}\\
    \mathbf{H}_{\text{M,R}}^{\mathsf{H}} &  \mathbb{T}(\bV)\\
    \end{bmatrix}
    \succeq 0.    
\end{equation}
The solution of the Toeplitz matrices $\mathbb{T}(\bC)$ and $\mathbb{T}(\bV)$ leads us to the recovery of the angles $\boldsymbol{\theta}_{\text{M,R}}$ and $\boldsymbol{\phi}_{\text{M,R}}$, respectively, by applying the root MUSIC algorithm~\cite{MUSIC}. We consider the order information of the angles and the number of paths as a priori information. We estimate the path gain vector $\boldsymbol{\rho}_{\text{M,R}}$ by applying the least squares (LS) method, which results in
\begin{equation}
    \label{eq_path_gain}
    \hat{\boldsymbol{\rho}}_{\text{M,R}} = 
   \Big[\beta_{\text{1}}(\mathbf{X}^{\mathsf{T}}\otimes \mathbf{W}_{\text{H}})\big((\bA^*(\hat{\boldsymbol{\theta}}_{\text{M,R}}) \odot \bA(\hat{\boldsymbol{\phi}}_{\text{M,R}})\big)\Big]^{\dagger}\mathbf{y}_{\text{H}},
\end{equation}
where $\mathbf{y}_{\text{H}} = \mathrm{vec}({\bY_{\text{H}}})$. 

\subsection{Stage 2 CE}
In the second stage, we target at the recovery of the remaining channel parameters from the received signal at BS. In order to facilitate the CE, we use the estimate of  $\hat{\mathbf{H}}_{\text{M,R}}$ from the first stage, which results in 
\begin{equation}
    \label{eq:hatU}
    \hat{\bU}=[\boldsymbol{\Omega}_{1}\hat{\mathbf{H}}_{\text{M,R}}\mathbf{X},\cdots,\boldsymbol{\Omega}_{K}\hat{\mathbf{H}}_{\text{M,R}}\mathbf{X}] \in \mathbb{C}^{N_{\text{R}}\times TK}.
\end{equation}
We formulate the atomic norm for the matrix $\bH_{\text{R,B}}$ as 
\begin{equation}
    \|\bH_{\text{R,B}}\|_{\mathcal{A}_{\text{N}}} = \inf \{q: \bH_{\text{R,B}} \in \mathrm{conv}({\mathcal{A}_{\text{N}}}) \} .
\end{equation}
By following the SDP formulation, we can write 
\begin{equation}
    \nonumber
    \|\mathbf{H}_{\text{R,B}}\|_{\mathcal{A}_{\text{N}}} = \mathrm{inf}_{\{\mathbf{S},\mathbf{O}\}} \Big\{\frac{1}{2 N_\text{R}} \mathrm{Tr}(\mathbb{T}(\mathbf{S})) + \frac{1}{2 N_\text{B}} \mathrm{Tr}(\mathbb{T}(\mathbf{O})) \Big\}
\end{equation}
\begin{center}
    \begin{equation}
        \textrm{s.t}   
        \begin{bmatrix}
        \mathbb{T}(\mathbf{S}) & \mathbf{H}_{\text{R,B}}\\
        \mathbf{H}_{\text{R,B}}^{\mathsf{H}} & \mathbb{T}(\mathbf{O})
    \end{bmatrix}
    \succeq 0.
    \end{equation}
\end{center}
where $\mathbb{T}(\mathbf{S})$ and $\mathbb{T}(\mathbf{O})$ are 2-level Toeplitz matrices. In order to recover the angles, we formulate the problem as 
\begin{equation}
    \hat{\bH}_{\text{R,B}}= \arg \min_{\bH_{\text{R,B}}} \nu \|\bH_{\text{R,B}}\|_{\mathcal{A}_{\text{N}}} + \frac{1}{2} \|\beta_{\text{2}} \mathbf{W}_{\text{B}}^{\mathsf{H}}\mathbf{H}_{\text{R,B}}\hat{\bU} - \bY\|^{2}_{\mathrm{F}},
\end{equation}
where $\nu \varpropto \sigma\sqrt{N_\text{B} N_\text{R}\log (N_\text{B} N_\text{R}})$
By following the SDP formulation, we expressed the problem as
\begin{equation}
            \nonumber
          \hat{\mathbf{H}}_{\text{R,B}} = \arg \min_{\bH_{\text{R,B}},\mathbf{{S}},\mathbf{{O}}} \frac{\nu}{2 N_\text{R}} \mathrm{Tr}(\mathbb{T}(\mathbf{S})) +  \frac{	\nu}{2 N_\text{B}} \mathrm{Tr}(\mathbb{T}(\mathbf{O})) 
\end{equation}
\begin{equation}
        \nonumber
        + \frac{1}{2} \|\beta_{\text{2}} \mathbf{W}_{\text{B}}^{\mathsf{H}}\mathbf{H}_{\text{R,B}}\hat{\bU} - \bY\|^{2}_{\mathrm{F}}
\end{equation}
\begin{equation}
    \textrm{s.t} \begin{bmatrix}
    \mathbb{T}(\mathbf{S}) & \mathbf{H}_{\text{R,B}}\\
    \mathbf{H}_{\text{R,B}}^{\mathsf{H}} &  \mathbb{T}(\mathbf{O})\\
    \end{bmatrix}
    \succeq 0,    
\end{equation}
The angles $\boldsymbol{\theta}_{\text{R,B}}$ and $\boldsymbol{\phi}_{\text{R,B}}$ can be recovered based on the solution of $\mathbb{T}(\mathbf{S})$ and $\mathbb{T}(\mathbf{O})$, respectively, via root MUSIC algorithm~\cite{MUSIC}. 
Similarly, the estimate of the path gain vector is addressed by LS, as
\begin{equation}
    \hat{\boldsymbol{\rho}}_{\text{R,B}} = 
   \Big[\beta_{\text{2}}(\hat{\mathbf{U}}^{\mathsf{T}}\otimes \mathbf{W}_{\text{B}}^{\mathsf{H}})\big((\bA^*(\hat{\boldsymbol{\theta}}_{\text{R,B}}) \odot \bA(\hat{\boldsymbol{\phi}}_{\text{R,B}})\big)\Big]^{\dagger}\mathbf{y},
\end{equation}
where $\mathbf{y} = \mathrm{vec}({\mathbf{Y}})$. Based on the estimation of the channel parameters, we calculate the angle differences and the products of path gains associated with the RIS for the purpose of comparison with the passive RIS benchmark~\cite{he2020anm}. The angle differences are functions of the estimates of $\boldsymbol{\phi}_{\text{M,R}}$ and $\boldsymbol{\theta}_{\text{R,B}}$, expressed as
\begin{align}
\label{eq_ang_diff}
    [\hat{\boldsymbol{\Delta}}]_{lp} &=  \mathrm{asin} \big[ \sin{([\hat{\boldsymbol{\phi}}_{\text{M,R}}]_{l})} - \sin{([\hat{\boldsymbol{\theta}}_{\text{R,B}}]_{p})} \big],\nonumber\\
    &\text{for}\; l = 1,\cdots, L_{\text{M,R}}, p = 1,\cdots, L_{\text{R,B}}.
\end{align}
Moreover, we also define the vectorization of $\hat{\boldsymbol{\Delta}}$ as $\hat{\boldsymbol{\delta}}=\text{vec}(\hat{\boldsymbol{\Delta}})$. The estimated products of path gains $\hat{\boldsymbol{\rho}} \in \C^{L_{\text{R,B}}L_{\text{M,R}} \times 1}$ are defined as follows: 
\begin{equation}
\label{eq_prod_path}
    \hat{\boldsymbol{\rho}} = \hat{\boldsymbol{\rho}}_\text{R,B} \otimes \hat{\boldsymbol{\rho}}_\text{M,R}.
\end{equation}
The training overhead for  hybrid RIS CE is 
\begin{equation}
    T_\text{H} =  K T \Big\lceil\frac{N_{\text{C,B}}}{N_{\text{RF,B}}}\Big\rceil
    \Big\lceil\frac{M}{N_{\text{RF,R}}}\Big\rceil,
\end{equation}
where $N_{\text{RF,B}}$ is the number of RF chains at BS. 

\subsection{Design of Phase Control Matrix and Beamforming Vectors}
For the sake of brevity, we briefly present the design of the phase control matrix and the beamforming vectors. We clarify that we follow the same procedure for both the hybrid and passive RIS architectures. 
We formulate the design of the phase control matrix in order to maximize the power (a.k.a. squared Frobenius norm) of $\bG$, defined in~\eqref{eq:g}. Moreover, the design of the active beamforming vectors is based on the reconstruction of the entire channel $\hat{\bH} = \hat{\bH}_{\text{R,B}}\hat{\boldsymbol{\Omega}}\hat{\bH}_{\text{M,R}}$. We conduct singular value decomposition (SVD) on $\hat{\bH}$ and use the left and right singular vectors associated with the largest singular value as BS and MS beamforming vectors, respectively. 

\subsection{Passive RIS CE via ANM}

In the first block of CE, i.e., $k =1$, we target at the extraction of the AoAs at BS and AoDs at MS, with $\mathbf{X}_1 \in \mathbb{C}^{N_{\text{M}}\times N_0}$, $\mathbf{W}_1 \in \mathbb{C}^{N_{\text{B}} \times M_0}$, and $\boldsymbol{\Omega}_1$. Based on the estimates $\{\hat{\boldsymbol{\theta}}_{\text{M,R}},\hat{\boldsymbol{\phi}}_{\text{R,B}} \}$ in the first stage, we design the sequential beam matrices and the combining matrices for the second stage, resulting in $\mathbf{X}_k = \bA(\hat{\boldsymbol{\theta}}_{\text{M,R}} ) \in \mathbb{C}^{N_{\text{M}}\times L_\text{M,R}} $ and $\mathbf{W}_k  = \bA(\hat{\boldsymbol{\phi}}_{\text{R,B}})\in \mathbb{C}^{N_{\text{B}} \times L_{\text{R,B}}}$, for $k =2,\cdots, K$. In the second stage, we target at the recovery of the angle differences and the products of path gains. Note that during this stage we keep the training matrix and the combining matrix fixed, while the RIS phase control matrix varies from block to block, i.e., $\mathbf{\Omega}_2 \neq \mathbf{\Omega}_3 \neq \cdots \neq \mathbf{\Omega}_K$. More details can be referred to~\cite{he2020anm}. After CE, the BS sends the estimates to the RIS controller and BS for the design of RIS phase control matrix and beamformer at MS for data transmission purpose.


\section{Performance Evaluation}
\label{section: perfomance}
\begin{figure} [!t]
    \centering
    \includegraphics[width=0.4\textwidth]{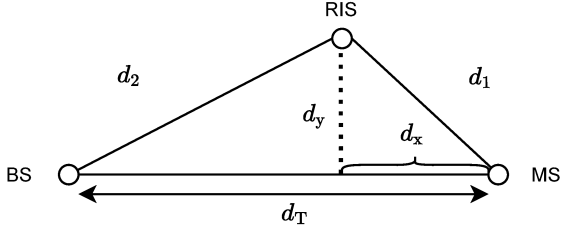}
    \caption{System setup. }
    \label{fig:path_loss}
    \vspace{-0.6cm}
\end{figure}

In this section, we describe the parameter setup and the performance metrics. The propagation path gains are distributed as $\mathcal{CN} (0,1)$. We set $N_{\text{B}} = 16$, $N_{\text{R}} = 32$,  $N_{\text{M}}= 16$, $L_{\text{R,B}} = L_{\text{M,R}} = 2$. 
We perform 1000 trials to average out the results. For simplicity, we assume 2-Dimensional Cartesian coordinate system. Fig.~\ref{fig:path_loss} shows the locations of the BS, RIS, and MS. We set the location of the BS, RIS and MS as~$(0,0)$, $(x,y)$, $(x_{T},0)$, respectively, where $x = d_{\text{T}} - d_\text{x}$, $y= d_\text{y}$ and $x_{T}=d_{\text{T}}$. The distance between MS and RIS is $d_{\text{1}} = {\sqrt{d_{\text{x}}^2 + d_{\text{y}}^2}}$, while the distance between BS and RIS is $d_{\text{2}} = {\sqrt{(d_{\text{T}} - d_{\text{x}})^2 + d_{\text{y}}^2}}$. 
The path loss as a function of distance is given by~\cite{zhang2020capacity}, $\beta(d_{\text{1}}) =\beta_\text{0}\Big(\frac{d_{0}}{d_{\text{1}}} \Big)^{\gamma}$ and $\beta(d_{\text{1}}, d_{\text{2}})=\beta_\text{0}\Big(\frac{d_{0}}{d_{\text{1}}d_{\text{2}}} \Big)^{\gamma}$,
where $\beta_\text{0}$ is the path loss with the reference distance $d_{\text{0}}$, defined as $\beta_\text{0} = (\frac{\lambda}{4 \pi d_{\text{0}}})^{2}$, $\lambda$ denotes the wavelength, $c$ is the speed of the light in meters per second ($c=3\times10^{8}$), and $f_{c}$ the carrier frequency in Hz. We set  $d_{0}$ as $1$ [m], the path loss exponent $\gamma=3$, $f_{c}=28$ MHz, the noise power density as $-173$ dBm/Hz, the bandwidth as $100$ MHz, and the transmit power as $\text{P}_{\text{t}} = 0:5:20$ dBm. 


To make the performance comparison fair, we keep the training overhead the same for both the hybrid RIS and the passive RIS. We evaluate the performance considering different number of active elements at RIS. Table~\ref{tab1:Setups} summarizes the parameter setup for the hybrid RIS architecture. 
\begin{table}
    \centering
    \caption{Parameter setup for the hybrid RIS architecture.}
    \label{tab1:Setups}
    \begin{tabular}{lccccccccc}
    \toprule
    \textbf  & {$N_{\text{R}}$}& {$M$} & {$N_{\text{RF,R}}$} & {$K$}  & {$T$} & {$N_{\text{C,B}}$} &  {$N_{\text{RF,B}}$} &{$T_{\text{H}}$} \\
   \midrule
        {Setup 1}  & 32 &$8$ & $8$ & $5$ & $8$ & $8$ & $8$ & $40$\\
        {Setup 2}  & 32 &$4$ & $4$ & $5$ & $8$ & $8$ & $8$ & $40$ \\
        {Setup 3}  & 32 &$2$ & $2$ & $5$ & $8$ & $8$ & $8$ & $40$ \\
   \bottomrule
    \end{tabular}
\end{table}
In order to guarantee the super-resolution estimation, we assume the spatial frequencies are separated by at least $\left(\frac{4}{N_{\text{B}}}\right)$,  $\left(\frac{4}{N_{\text{R}}}\right)$, $\left(\frac{4}{N_{\text{M}}}\right)$. 

\begin{figure} [h!]
    \centering
    \includegraphics[width = \columnwidth]{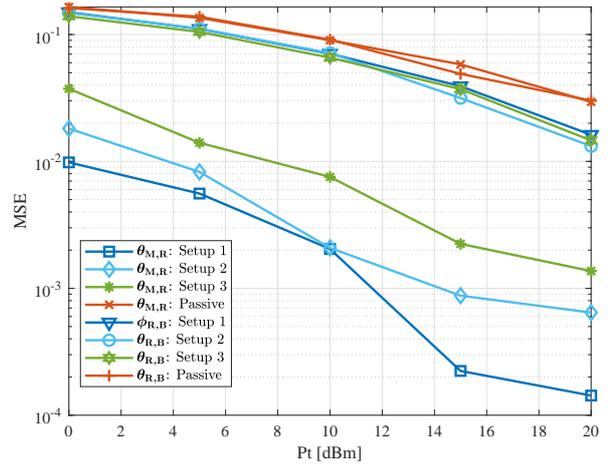}
    \caption{MSE of the angular parameters.}
    \label{fig:results2}
    \vspace{-0.6cm}
\end{figure}

\subsection{Performance Metrics}
We evaluate the performance in terms of the MSE\footnote{The MSE can be formulated by considering the sine of the angles, i.e., spatial frequencies. We clarify that the results  will be consistent with these based on estimated angular parameters. } of the estimates of AoDs, AoAs, angle differences and the products of path gains. Also, we evaluate the performance in terms of average SE in (bits/s/Hz).
\subsection{Simulation Results}
The simulation results for the estimation of channel parameters $\boldsymbol{\theta}_{\text{M,R}}$ and $\boldsymbol{\phi}_{\text{R,B}}$ are shown in Fig.~\ref{fig:results2}, where we consider $d_{\text{T}}=25$ [m], $d_{\text{x}}= 10$ [m], $d_{\text{y}}=2$ [m]. We compare our results with the benchmark scheme, i.e., the passive RIS detailed in~\cite{he2020anm}. For the estimation of $\boldsymbol{\theta}_{\text{M,R}}$, the setups 1--3 of the hybrid RIS outperform the passive RIS. The performance of the setup 1 brings the best performance among the hybrid setups, while the setup 3 has the lowest power consumption due to the least number of active elements and RF chains at the RIS. The results can be explained by the effect of path loss on the RIS CE. In our proposed method, we can perform CE at the RIS, where the path loss is proportional to the distance $d_{1}$. On the other hand, when the CE is performed at BS, the path loss is proportional to $d_{1}d_{2}$, which degrades the performance.

\begin{figure}[!t]
    \centering
    \includegraphics[width = \columnwidth]{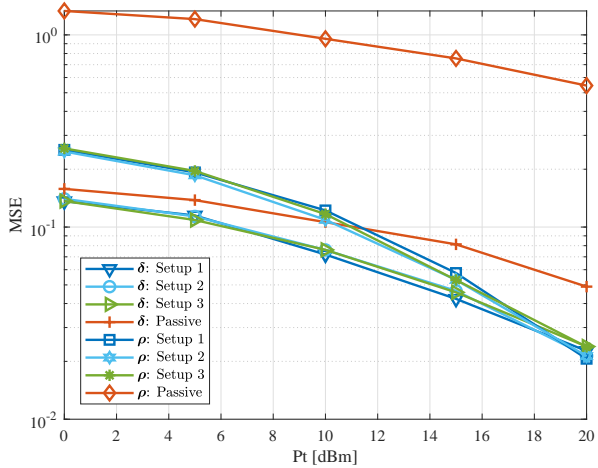}
    \caption{MSE of the products of path gains and angle differences.}
    \label{fig:MAE}
    \vspace{-0.4cm}
\end{figure}

Regarding the MSE of $\boldsymbol{\delta}$ and  $\boldsymbol{\rho}$, the hybrid RIS also has better performance than the passive RIS. Also, we can see in Fig.~\ref{fig:MAE} that the performance of hybrid setups are quite similar. The estimates of the coupled parameters $\boldsymbol{\delta}$ and $\boldsymbol{\rho}$ depend on the estimates of $\boldsymbol{\phi}_{\text{M,R}}$ and $\boldsymbol{\rho}_{\text{M,R}}$ in the first stage and these of $\boldsymbol{\theta}_{\text{R,B}}$ and $\boldsymbol{\rho}_{\text{R,B}}$ in the second stage. In the first CE stage, we can obtain a better estimate of the channel parameters due to the lower path loss. However, in the second CE stage, the path loss is more severe, which results in a poor estimation of $\boldsymbol{\theta}_{\text{R,B}}$ and $\boldsymbol{\rho}_{\text{R,B}}$. The poor estimation in the second stage will dominate the resultant performance in Figs.~\ref{fig:MAE}.   

\begin{figure}[!t]
    \centering
    \includegraphics[width = \columnwidth]{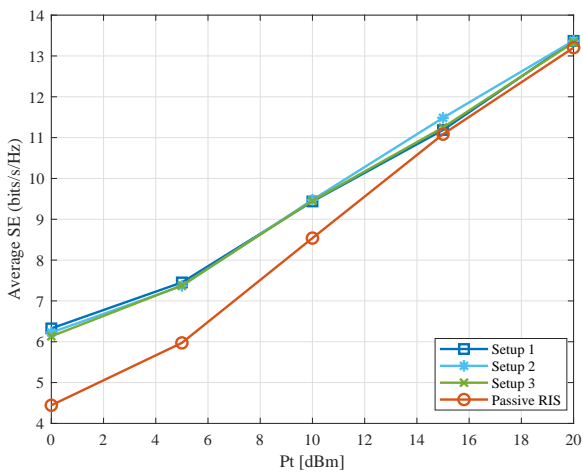}
    \caption{Average SE.}
    \label{fig:se}
    \vspace{-0.6cm}
\end{figure}

Fig.~\ref{fig:se} shows the average SE of our proposed method and the passive RIS~\cite{he2020anm}. The setups 1--3 of our proposed method can achieve better performance than the passive RIS for low transmit power, for example $P_{\text{t}} = 5 $ dBm. However, for high transmit power, $P_{\text{t}} = 20$ dBm, the hybrid and the passive RIS can achieve similar results. Also, we can obtain similar performances even with reduced number of active elements. Since the active elements are proposed only for collecting received signals, their position does not affect the CE. Moreover, the position of the RIS can bring benefits for the hybrid RIS. Since we can perform CE at the RIS, we can obtain better performance when the RIS is closer to the MS.

\section{Conclusions}
\label{section: conclusions}
In this paper, we have studied the channel estimation in hybrid RIS aided mmWave MIMO systems and proposed a two-stage CE scheme via ANM. The results have shown that our proposed CE method could bring better performance compared to the passive RIS. Nevertheless, it is worth mentioning that due to presence of the active elements, the power consumption is higher. As our future work, we will investigate  the effect of the position of the RIS on CE as well as SE, affected by the semi-passive beamforming at the RIS.

\section{Acknowledgements}
This work has been financially supported in part by the Academy of Finland (ROHM project, grant 319485), European Union's Horizon 2020 Framework Programme for Research and Innovation (ARIADNE project, under grant agreement no. 871464), and Academy of Finland 6Genesis Flagship (grant 318927).

\bibliographystyle{IEEEtran}
\bibliography{IEEEabrv,references}

\end{document}